\documentclass[sigconf]{acmart}
\setcopyright{none}
\renewcommand\footnotetextcopyrightpermission[1]{}

\AtBeginDocument{%
  }

\acmConference[]{}{}{}

\usepackage[inkscapelatex=false]{svg}
\usepackage{multirow}
\usepackage{graphicx}
\usepackage[table,xcdraw]{xcolor}
\usepackage{booktabs}

\begin{document}

\title{SPPAM: \textbf{S}ignature \textbf{P}attern \textbf{P}rediction and \textbf{A}ccess-\textbf{M}ap Prefetcher }

\author{Maccoy Merrell, Lei Wang, Stavros Kalafatis, Paul V. Gratz}
\email{{ maccoy.merrell, wilsonwang2019, skalafatis-tamu, pgratz}@tamu.edu}
\orcid{0009-0000-5161-2013, 0009-0007-0273-236X, 0000-0002-7543-2479, 0000-0001-7120-7189}
\affiliation{%
  \institution{Texas A\&M University}
  \city{College Station}
  \state{Texas}
  \country{USA}
}

\begin{abstract}
 The discrepancy between processor speed and memory system performance continues to limit the performance of many workloads. To address the issue, one effective and well studied technique is cache prefetching. Many prefetching designs have been proposed, with varying approaches and effectiveness.  For example, SPP is a popular prefetcher that leverages confidence throttled recursion to speculate on the future path of the program's references, however it is very susceptible to the reference reordering of higher-level caches and the out-of-order core.  Orthogonally, AMPM is another popular approach to prefetching which uses reordering-resistant access maps to identify patterns within a region but is unable to speculate beyond that region.  In this paper, we propose SPPAM, a new approach to prefetching, inspired by prior works such as SPP and AMPM, while addressing their limitations. SPPAM utilizes online-learning to build a set of access-map patterns. These patterns are used in a speculative lookahead which is throttled by a confidence metric. Targeting the second-level cache, SPPAM alongside state-of-the-art prefetchers Berti and Bingo improves system performance by 31.4\% over no prefetching and 6.2\% over the baseline of Berti and Pythia.
\end{abstract}

\maketitle

\pagestyle{empty}

\section{Introduction}
As is well known, the processor has outpaced the speed of the memory system and memory system performance
has become a major bottleneck. This phenomenon, known as the \textit{Memory Wall}, plagues memory-bound
workloads. A common and effective technique to address this issue is hardware prefetching which
speculatively populates the caches with data, ideally maximizing coverage, timeliness and accuracy \cite{taxnomy}. Many prefetcher designs have been proposed, among these being
AMPM \cite{AMPM}, Bingo \cite{Bingo}, Pythia \cite{Pythia}, Berti \cite{Berti} and SPP \cite{SPP}. Each of these prefetchers has certain strengths and weaknesses, targeting different cache levels
and offering widely varying state overheads.  SPP~\cite{SPP} is a popular prefetcher that leverages confidence throttled recursion to speculate on the future path of the program's references.  SPP, however, requires consistent replays of the same pattern of access, making it very susceptible to the reference reordering caused by higher-level caches and the OoO core.  Orthogonally, AMPM, though somewhat older, has been implemented in recent systems~\cite{ampm_patent,Bruce_2023}.  AMPM uses reordering-resistant reference pattern matching via access maps to identify patterns within a region.  While this technique speculates well within the given region, it does not extend patterns beyond that space.

 In this paper, we propose SPPAM, a new approach to prefetching, inspired by prior works such as SPP and AMPM, while addressing their limitations. SPPAM utilizes online-learning to build a set of access-map patterns. These patterns are used in a speculative lookahead which is throttled by a confidence metric. Targeting the second-level cache, SPPAM alongside state-of-the-art prefetchers Berti and Bingo improves system performance by 31.4\% over no prefetching and 6.2\% over the baseline of Berti and Pythia.

\section{Previous Works}

Second-level-cache (L2C) prefetching offers the most potential compared to other cache levels \cite{R-Max}. The size of the cache and its private nature allow for large working sets and speculative data to coexist, which cannot be said of the smaller first-level data caches (L1D). Unlike the last-level-cache (LLC), the L2C is close to the core and has lower access latency which means successful prefetches are more effective at reducing apparent latency from the CPU. It also offers challenges, primarily in the form of information. The L2C receives accesses that are heavily reordered by the core, and these reordered requests are heavily filtered and randomly decimated by the L1D. Learning from these patterns is difficult without complex pattern recognition techniques, and such techniques are prone to over-differentiation.

The Signature Path Prefetcher (SPP) addresses some of the problems of L2C prefetching by identifying repeated access patterns in the form of delta signatures. These signatures are page-, offset-, and time-agnostic, instead only recording the past three differences between consecutive block accesses in a page and predicting the subsequent one through previous access patterns. This allows for SPP to aggressively collect and apply data patterns in prefetching even in data regions with no history, with high accuracy. These predictions can be used recursively by appending the predicted delta to the signature, allowing SPP to speculate deeply into an access stream upon only a small number of initial accesses. Over time, this history leads to emergent prefetching behavior allowing SPP to predict complex access patterns. This technique has a clear weak point: its signatures are order-dependent. If requests to the L2C are reordered (likely, as previously described) then the signature will change, resulting in the same memory pattern being reclassified upon subsequent accesses. This prevents some patterns from being correctly identified and its state from being efficiently used, thus causing losses in performance.

A popular order-agnostic prefetcher design is known as AMPM. AMPM, unlike SPP, checks the use of nearby blocks in a region rather than deltas. The pattern of block accesses can then be used to inform which other blocks need to be pulled into the cache. AMPM's algorithm is resilient to reordering interference, as these access maps encode no order information. AMPM typically relies on static pattern matching, meaning that access patterns that are outside AMPM's internal catalog are ignored. Likewise, AMPM can make only a single set of predictions from a pattern, solely within the scope of the pattern itself. Workloads whose following accesses do not exactly match AMPM's pre-programmed expectations yield many incorrect predictions and thus performance loss. Other workloads may be inherently incompatible with AMPM, particularly if they have access patterns with significant internal page fragmentation. Techniques like scanning can help compensate for this, where areas around the initial access are evaluated for potential pattern matches, but if too many regions are active at once AMPM begins losing region history and thus suffers substantial performance degradation.
These two strategies are effective in isolation despite their flaws, suggesting that future spatial data prefetchers that which to supersede these designs should leverage their strengths and avoid these weaknesses.

\section{Motivation}
Utilizing history for making predictions is a powerful technique utilized not only in hardware prefetching but also in branch prediction. Unlike branch prediction, identifying the history of memory accesses requires far more discernment. AMPM, SPP, and other spatial data prefetchers correctly identified that segmenting the histories of memory accesses by region leads to many patterns being found that are not readily apparent from observing the access stream in-order. Crucially, data structures are not merely read, but operated upon. These data structures may be accessed in a well-ordered and observable way, but this behavior may be intermittent with other memory being utilized in-between. Even when these patterns should be easily identified, processors issue memory requests out-of-order and some memory requests may hit in upper-level caches while others will not. These patterns fluctuate under short-term observation, leading prefetchers who attempt to learn detailed delta-based histories to fail.
Instead of delta-based prefetching, a prefetcher which encodes a history of which nearby blocks have been utilized (such as AMPM) is immune to this fluctuation. While AMPM does not learn from its history, a different prefetcher utilizing this strategy could delay learning any current patterns until it is certain the history has stabilized, and likewise can revisit old history and re-evaluate it as patterns emerge.

\section{Design Overview}
We propose the Signature Pattern Prediction and Access Map (SPPAM) prefetcher, which combines novel techniques as well as those from SPP and AMPM into a single system, attempting to rectify each strategies' individual weaknesses. We utilize AMPM's region-tracking technique to record accesses to individual regions in access maps. For each of these accesses, the segment of the access map before the current access can be extracted and used as a pattern. A pattern can be taken from both before (positive) and after (negative) the offset into the region. These patterns are then used to access a direct-mapped pair of positive and negative pattern tables. These tables track the most-common access map pattern pairs, with a fully-associative set of matches being stored within each entry as a prediction. The most-common entry is selected for use as the prediction, which indicates which of the next consecutive blocks in the region will be prefetched.
The predicted sequence is of the same width as the index into the pattern table, so it can be used recursively to generate prefetches deeper into the region, similarly to SPP's look-ahead feature, but far denser allowing for many predictions for each table lookup.
Along with these improvements, SPPAM incorporates several different throttling techniques, taking in feedback in the form of DRAM bandwidth consumption and prefetch utilization.

\begin{figure}[]
	\centering
	\includesvg[width=0.70\columnwidth]{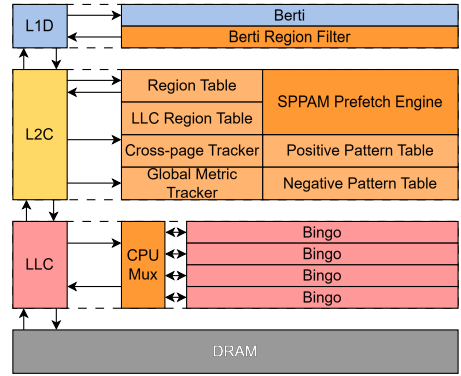}
        \caption{SPPAM Architecture}
        \label{fig:sppam_overview}
\end{figure}

Figure~\ref{fig:sppam_overview} shows SPPAM in the L2C, leaving both the L1D and LLC available for other prefetcher designs. For our design, we utilize Berti as the L1D prefetcher and Bingo as the LLC prefetcher. Minor improvements in these systems were made, in the figure these additions are shown in orange, with Berti utilizing an improved filter inspired by AMPM's region tables and Bingo being core-partitioned, preventing cross-core interference in Bingo's learning. Berti also provides region-association information to SPPAM from the virtual space, allowing SPPAM to use cross-page information.

\subsection{Region Table and Pattern Scraping}
\begin{figure}[]
	\centering
	\includesvg[width=0.9\columnwidth]{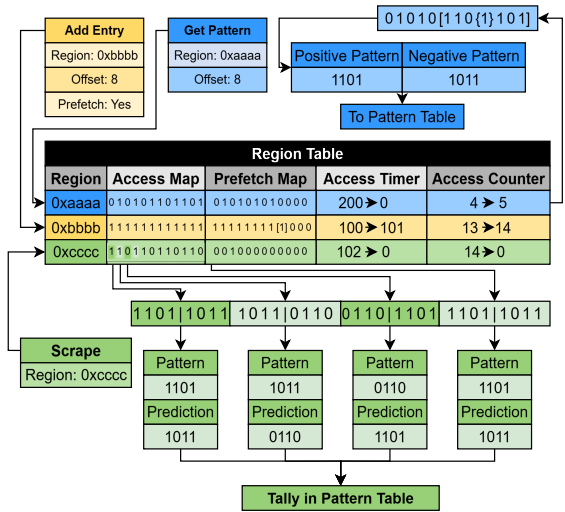}
        \caption{Region Table Architecture}
        \label{fig:region_access}
\end{figure}

SPPAM's region table is managed as a set-associative table indexed by region number. Within each region table entry
bitmaps are stored both for the accesses into the region and prefetches done into the region. The former bitmap is used to identify patterns, while the latter is used as a filter - being cleared and set as blocks in the region are populated into the cache.
Region sizing determines the size of the bitmap and overall efficiency of representing the full memory space. Smaller region sizes introduce more state overhead, while larger sizes can potentially lead to significant internal fragmentation and wasted space within the region's bitmaps. We found that sizing regions to 4KiB is a good compromise, preventing additional complexity from sequencing non-consecutive physical pages into regions while not being so small as to incur significant storage overheads for workloads with large memory spaces.
Figure~\ref{fig:region_access} shows different access types into the region table. Either additional bits are added or removed from the internal bitmap structures, or patterns are retrieved from those bitmaps for use in the pattern table.
Two additional pieces of data are stored within each region table entry compared to AMPM. These exist to facilitate a SPPAM-specific feature - pattern scraping. First, a timer has been added that indicates how long the region has gone without access. The second is an access counter which stores how many recent accesses have occurred into the region. If either the timer or the counter exceeds their respective thresholds, a region scrape is initiated for the region.
Region scraping is how SPPAM builds its pattern tables. Once a region has had sufficient activity or falls out-of-scope, the access map must be processed to identify which access patterns were present. The access map is navigated both from front to back and back to front for the positive and negative pattern tables. A rolling window of the access map (sized according to the pattern table) and the following pattern (offset ahead/behind) is tallied into the pattern table, modifying the confidence of the predictions for the given pattern.

\subsection{Pattern Tables and Lookahead}
The pattern tables are direct-mapped and sized according to the pattern window used by SPPAM. Each pattern is used to index into the table and produce the predicted pattern that follows. Each entry within the pattern table includes an associative set of predictions. This technique assumes that many different predictions may be associated with a single pattern. The most common of these is expected to be the most useful, since it offers the most opportunities to prefetch. For this reason, only the most-common prediction is used. These tables are susceptible to interference - particularly from other predictions within a hamming-distance of 1 from the most-common pattern. The impact of this interference can be somewhat negated by properly timing region scraping to avoid incomplete patterns from being learned and thus reducing the chance of small variations of the same pattern being encoded.

\begin{figure}[]
	\centering
	\includesvg[width=0.95\columnwidth]{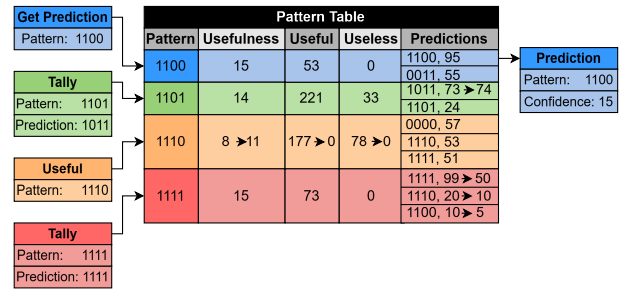}
        \caption{Pattern Table Architecture}
        \label{fig:pattern_table}
\end{figure}

Three different types of information are stored within the pattern table, with the overall architecture shown in Figure~\ref{fig:pattern_table}.
The prediction table tracks the confidence of different predictions. Confidence is increased each time a region scrape results in the predicted pattern being incremented, and the confidences of all entries in the table are halved when the confidence max is reached for any given pattern. The most-confident pattern above the minimum confidence is used as the prediction, with no valid pattern being returned if no prediction is above the minimum confidence. The least-confident pattern is chosen as the eviction candidate.
In addition to the prediction table, each entry also includes useful and useless counters. As prefetches are utilized or evicted from the cache, their corresponding region provides the patterns that predicted the prefetch and inform the pattern table. The patterns within the pattern table track the counts and when the total of useless and useful prefetches reaches a sampling threshold, the usefulness of the pattern is estimated.

\begin{figure}[]
	\centering
	\includesvg[width=0.8\columnwidth]{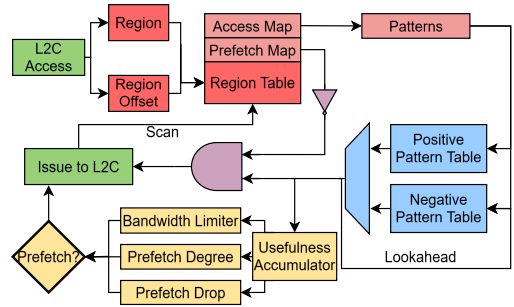}
        \caption{SPPAM Lookahead}
        \label{fig:lookahead}
\end{figure}

Figure~\ref{fig:lookahead} shows an overview of SPPAM's logic, including lookahead. Predicted patterns can be fed back into SPPAM to continue prefetching beyond the initial region offset. This can continue as long as the predicted pattern does not equal zero, at which point SPPAM proceeds with the standard AMPM strategy of scanning outwards from the original region offset to find more prefetch candidates. Upon each subsequent lookahead, SPPAM reduces the usefulness proportional to the predicted usefulness of the new pattern. This is continued until the prefetch degree is satisfied or usefulness falls too low to continue. How these thresholds are chosen is discussed further in the following section.

\subsection{Global Metrics and Bandwidth}
Alongside pattern usefulness, global usefulness is also estimated using a sampling procedure. Once enough samples of useful and useless
prefetches are accumulated, the 4-bit global usefulness is set and remains until the next sampling.
Pattern usefulness is effective at estimating the usefulness of a pattern as long as feedback from useful and useless prefetches can
return in time for the region to still reside in the region table, otherwise global usefulness must be used.
To determine if this is the case, a 4-bit region lifespan metric is tracked by SPPAM. Region lifespan is the proportion of useless prefetches evicted from the cache whose regions are still within the region table. A low region lifespan indicates that pattern usefulness is unreliable and SPPAM will instead use global usefulness.
Usefulness controls two throttling mechanisms within SPPAM. The first of these is the prefetch degree, the number of prefetches that may be issued from a single cache access. As usefulness drops, the prefetch degree drops. Because of SPPAM's (and for that matter AMPM's) nature, this only reduces the rate at which prefetches are issued, and will not prevent a prefetch from eventually being issued.
The second mechanism is prefetch dropping. If SPPAM predicts low usefulness for a given prefetch, then it has a chance of dropping the prefetch. The rate at which this is done increases as usefulness decreases. SPPAM never drops all prefetches, as samples are still required to determine usefulness. Dropping is implemented using a 7-bit counter that tracks the current cycle count, with prefetches being dropped if the cycle count that triggered the prefetch was below a certain value.
A third throttling mechanism is controlled by available DRAM bandwidth, a 4-bit value provided by the memory controller.
SPPAM reduces the prefetch degree by an amount proportional to the consumed bandwidth.
The prefetch degree, prefetch drop chance, and bandwidth throttling are all hard-coded as 16-entry tables, indexed by
the 4-bit usefulness and 4-bit bandwidth utilization respectively.

\subsection{Prefetch Filtering and LLC Prefetching}
SPPAM filters prefetches via its region table. If a prefetch has already been issued, it will not be issued again until
the region is out of scope or that prefetch is evicted from the cache.
SPPAM may choose to fill only into the LLC if the MSHR of the L2C is currently fully occupied. The region table cannot track the occupancy of blocks within the LLC, so SPPAM utilizes a second region table that tracks only recently accessed or prefetched blocks within the LLC. If a block is bound for the LLC and resides in the LLC region table, it is dropped and not used. The LLC region table clears its entries as matching blocks in the L2C are evicted.

\begin{table}[]
\resizebox{0.90\columnwidth}{!}{%
\begin{tabular}{|l|l|l|}
\hline
\rowcolor[HTML]{000000} 
{\color[HTML]{FFFFFF} \textbf{Prefetcher}}                                       & {\color[HTML]{FFFFFF} \textbf{Description}}                                                           & {\color[HTML]{FFFFFF} \textbf{Total State}}                                                                                            \\ \hline
\rowcolor[HTML]{656565} 
\cellcolor[HTML]{656565}{\color[HTML]{FFFFFF} }                                  & \cellcolor[HTML]{C0C0C0}Delta Table: 64 entries, 16 deltas each: 2.54 KiB                             & \cellcolor[HTML]{656565}{\color[HTML]{FFFFFF} }                                                                                        \\ \cline{2-2}
\rowcolor[HTML]{656565} 
\cellcolor[HTML]{656565}{\color[HTML]{FFFFFF} }                                  & \cellcolor[HTML]{EFEFEF}History Table: 8 sets, 16 ways: 0.74 KiB                                      & \cellcolor[HTML]{656565}{\color[HTML]{FFFFFF} }                                                                                        \\ \cline{2-2}
\rowcolor[HTML]{656565} 
\cellcolor[HTML]{656565}{\color[HTML]{FFFFFF} }                                  & \cellcolor[HTML]{C0C0C0}L1D: 12-bit latency per line: 1.13 KiB                                        & \cellcolor[HTML]{656565}{\color[HTML]{FFFFFF} }                                                                                        \\ \cline{2-2}
\rowcolor[HTML]{656565} 
\cellcolor[HTML]{656565}{\color[HTML]{FFFFFF} }                                  & \cellcolor[HTML]{EFEFEF}Region Filter: 64 sets, 12 ways, 30-bit region tag, 64-bit access map, 4-bit, & \cellcolor[HTML]{656565}{\color[HTML]{FFFFFF} }                                                                                        \\
\rowcolor[HTML]{656565} 
\multirow{-5}{*}{\cellcolor[HTML]{656565}{\color[HTML]{FFFFFF} \textbf{Berti}}}  & \cellcolor[HTML]{EFEFEF}4-bit LRU: 9.19 KiB                                                           & \multirow{-5}{*}{\cellcolor[HTML]{656565}{\color[HTML]{FFFFFF} \textbf{13.60 KiB}}}                                                    \\ \hline
\rowcolor[HTML]{9B9B9B} 
\cellcolor[HTML]{9B9B9B}{\color[HTML]{FFFFFF} }                                  & \cellcolor[HTML]{C0C0C0}Pattern History Table: 512 sets, 15 ways: 56.25 KiB                           & \cellcolor[HTML]{9B9B9B}{\color[HTML]{FFFFFF} }                                                                                        \\ \cline{2-2}
\rowcolor[HTML]{9B9B9B} 
\cellcolor[HTML]{9B9B9B}{\color[HTML]{FFFFFF} }                                  & \cellcolor[HTML]{EFEFEF}Filter Table: 4 sets, 16 ways: 0.48 KiB                                       & \cellcolor[HTML]{9B9B9B}{\color[HTML]{FFFFFF} }                                                                                        \\ \cline{2-2}
\rowcolor[HTML]{9B9B9B} 
\cellcolor[HTML]{9B9B9B}{\color[HTML]{FFFFFF} }                                  & \cellcolor[HTML]{C0C0C0}Accumulation Table: 8 sets, 16 ways: 1.44 KiB                                 & \cellcolor[HTML]{9B9B9B}{\color[HTML]{FFFFFF} }                                                                                        \\ \cline{2-2}
\rowcolor[HTML]{9B9B9B} 
\multirow{-4}{*}{\cellcolor[HTML]{9B9B9B}{\color[HTML]{FFFFFF} \textbf{Bingo}}}  & \cellcolor[HTML]{EFEFEF}Prefetch Streamer: 8 sets, 16 ways: 1.86 KiB                                  & \multirow{-4}{*}{\cellcolor[HTML]{9B9B9B}{\color[HTML]{FFFFFF} \textbf{\begin{tabular}[c]{@{}l@{}}60.03 KiB\\ per core\end{tabular}}}} \\ \hline
\rowcolor[HTML]{656565} 
\cellcolor[HTML]{656565}{\color[HTML]{FFFFFF} }                                  & \cellcolor[HTML]{C0C0C0}Region Table: 128 sets, 24 ways, 29-bit region tag,                           & \cellcolor[HTML]{656565}{\color[HTML]{FFFFFF} }                                                                                        \\
\rowcolor[HTML]{656565} 
\cellcolor[HTML]{656565}{\color[HTML]{FFFFFF} }                                  & \cellcolor[HTML]{C0C0C0}72-bit region adjacency tag, 64-bit access map, 64-bit prefetch map,          & \cellcolor[HTML]{656565}{\color[HTML]{FFFFFF} }                                                                                        \\
\rowcolor[HTML]{656565} 
\cellcolor[HTML]{656565}{\color[HTML]{FFFFFF} }                                  & \cellcolor[HTML]{C0C0C0}4-bit access counter, 9-bit access timer, 5-bit LRU: 92.63 KiB                & \cellcolor[HTML]{656565}{\color[HTML]{FFFFFF} }                                                                                        \\ \cline{2-2}
\rowcolor[HTML]{656565} 
\cellcolor[HTML]{656565}{\color[HTML]{FFFFFF} }                                  & \cellcolor[HTML]{EFEFEF}LLC Region Table: 64 sets, 16 ways, 30-bit region tag, 64-bit access map,     & \cellcolor[HTML]{656565}{\color[HTML]{FFFFFF} }                                                                                        \\
\rowcolor[HTML]{656565} 
\cellcolor[HTML]{656565}{\color[HTML]{FFFFFF} }                                  & \cellcolor[HTML]{EFEFEF}4-bit LRU: 12.25 KiB                                                          & \cellcolor[HTML]{656565}{\color[HTML]{FFFFFF} }                                                                                        \\ \cline{2-2}
\rowcolor[HTML]{656565} 
\cellcolor[HTML]{656565}{\color[HTML]{FFFFFF} }                                  & \cellcolor[HTML]{C0C0C0}Pattern Table: 64 entries, 8-bit useful counter, 8-bit useless counter,       & \cellcolor[HTML]{656565}{\color[HTML]{FFFFFF} }                                                                                        \\
\rowcolor[HTML]{656565} 
\cellcolor[HTML]{656565}{\color[HTML]{FFFFFF} }                                  & \cellcolor[HTML]{C0C0C0}4-bit usefulness, 16 13-bit predictions, x2: 3.56 KiB                            & \cellcolor[HTML]{656565}{\color[HTML]{FFFFFF} }                                                                                        \\ \cline{2-2}
\rowcolor[HTML]{656565} 
\cellcolor[HTML]{656565}{\color[HTML]{FFFFFF} }                                  & \cellcolor[HTML]{EFEFEF}Cross-Page Tracker: 128 entries, 8-bit stream id, 36-bit region id,           & \cellcolor[HTML]{656565}{\color[HTML]{FFFFFF} }                                                                                        \\
\rowcolor[HTML]{656565} 
\cellcolor[HTML]{656565}{\color[HTML]{FFFFFF} }                                  & \cellcolor[HTML]{EFEFEF}1-bit direction: 0.71 KiB                                                     & \cellcolor[HTML]{656565}{\color[HTML]{FFFFFF} }                                                                                        \\ \cline{2-2}
\rowcolor[HTML]{656565} 
\cellcolor[HTML]{656565}{\color[HTML]{FFFFFF} }                                  & \cellcolor[HTML]{C0C0C0}Global Usefulness Tracker: 4-bit usefulness, 10-bit useful counter,           & \cellcolor[HTML]{656565}{\color[HTML]{FFFFFF} }                                                                                        \\
\rowcolor[HTML]{656565} 
\cellcolor[HTML]{656565}{\color[HTML]{FFFFFF} }                                  & \cellcolor[HTML]{C0C0C0}10-bit useless counter: 3 B                                                   & \cellcolor[HTML]{656565}{\color[HTML]{FFFFFF} }                                                                                        \\ \cline{2-2}
\rowcolor[HTML]{656565} 
\cellcolor[HTML]{656565}{\color[HTML]{FFFFFF} }                                  & \cellcolor[HTML]{EFEFEF}Region Lifespan Tracker: 4-bit lifespan, 8-bit hit counter,                   & \cellcolor[HTML]{656565}{\color[HTML]{FFFFFF} }                                                                                        \\
\rowcolor[HTML]{656565} 
\multirow{-13}{*}{\cellcolor[HTML]{656565}{\color[HTML]{FFFFFF} \textbf{SPPAM}}} & \cellcolor[HTML]{EFEFEF}8-bit miss counter: 2.5 B                                                     & \multirow{-13}{*}{\cellcolor[HTML]{656565}{\color[HTML]{FFFFFF} \textbf{109.16 KiB}}}                                                  \\ \hline
\end{tabular}%
}
\caption{State Overheads}
\label{tab:sppam_state}
\end{table}

\begin{table}[]
\resizebox{0.9\columnwidth}{!}{%
\begin{tabular}{|
>{\columncolor[HTML]{656565}}l 
>{\columncolor[HTML]{9B9B9B}}l |}
\hline
\multicolumn{2}{|l|}{\cellcolor[HTML]{000000}{\color[HTML]{FFFFFF} \textbf{SPPAM Configuration}}}                                                                                                                \\ \hline
\multicolumn{1}{|l|}{\cellcolor[HTML]{656565}{\color[HTML]{FFFFFF} Pattern Size: 6-bit}}            & {\color[HTML]{FFFFFF} Scrape Time Threshold: 1000}                                                         \\ \hline
\multicolumn{1}{|l|}{\cellcolor[HTML]{9B9B9B}{\color[HTML]{FFFFFF} Scrape Access Threshold: 14}}    & \cellcolor[HTML]{656565}{\color[HTML]{FFFFFF} Prefetch Degrees: {[}1,1,2,2,2,3,3,3,4,4,8,8,12,12,16,16{]}} \\ \hline
\multicolumn{1}{|l|}{\cellcolor[HTML]{656565}{\color[HTML]{FFFFFF} Lookahead Decay, Cutoff: 13, 7}} & {\color[HTML]{FFFFFF} Drop Chance: {[}123,120,110,110,80,50,10,0,0,0,0,0,0,0,0,0{]}}                       \\ \hline
\multicolumn{1}{|l|}{\cellcolor[HTML]{9B9B9B}{\color[HTML]{FFFFFF} Scan Distance: 16}}              & \cellcolor[HTML]{656565}{\color[HTML]{FFFFFF} Bandwidth Reduction: {[}0,0,0,0,1,1,1,1,2,2,2,4,4,4,8,8{]}}  \\ \hline
\multicolumn{1}{|l|}{\cellcolor[HTML]{656565}{\color[HTML]{FFFFFF} Prefetch Drop Cutoff: 8}}        & {\color[HTML]{FFFFFF} Region Lifespan Cutoff: 7}                                                           \\ \hline
\end{tabular}%
}
\caption{SPPAM Configuration}
\label{tab:sppam_config}
\end{table}

\begin{figure*}[]
	\centering
	\includesvg[width=1.9\columnwidth]{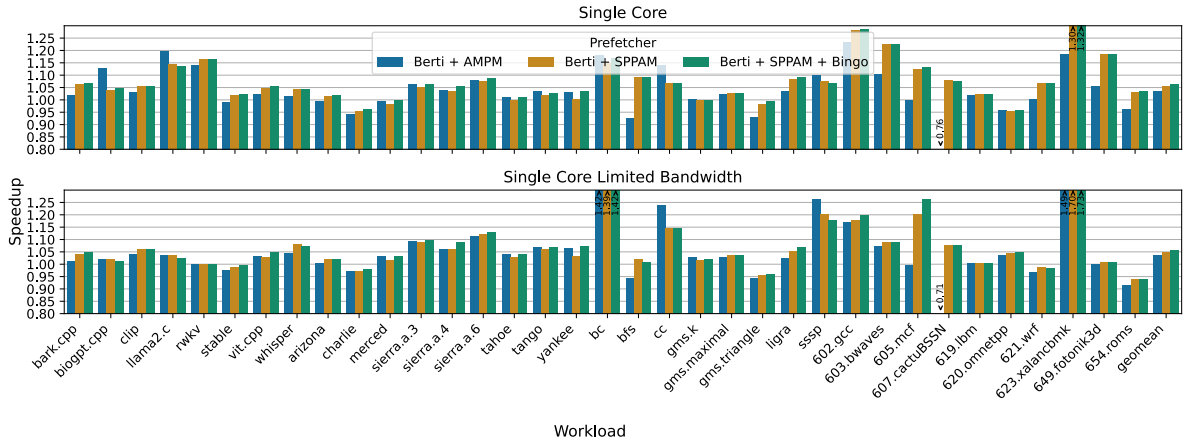}
        \caption{Single core speedup over Berti + Pythia across DPC4 traces}
        \label{fig:singlecore_speedup}
\end{figure*}

\subsection{Berti and Virtual Prefetching}
Berti is a state-of-the-art L1D virtual prefetcher and thus crosses page boundaries when prefetching. SPPAM operates within the physical space and thus cannot know which pages are consecutive. We modify Berti to pass along this information via a stream identification tag. This tag indicates that a sequence of prefetches is from the same access stream and the direction of the access stream. SPPAM utilizes this along with a small page crossing table to identify which regions are consecutive in both the forwards and backwards directions.
SPPAM can then "shadow" region edges, allowing patterns that cross page boundaries to both
be read and scraped. This also allows SPPAM to prefetch into new regions, utilizing the shadow-region information to jump-start the prefetching process. SPPAM does not issue prefetches across page boundaries, as it offers little to no performance benefit and presents potential security issues.

\begin{figure}[]
	\centering
	\includesvg[width=0.9\columnwidth]{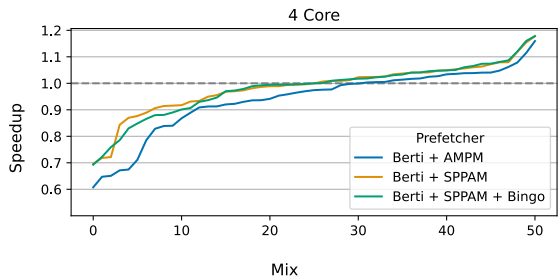}
        \caption{Sorted 4-core speedup over Berti + Pythia across 50 mixes}
        \label{fig:4core_speedup}
\end{figure}

\begin{table}[]
\resizebox{0.8\columnwidth}{!}{%
\begin{tabular}{|ll|}
\hline
\rowcolor[HTML]{000000} 
\multicolumn{2}{|l|}{\cellcolor[HTML]{000000}{\color[HTML]{FFFFFF} \textbf{System Configuration}}}                                                                                               \\ \hline
\rowcolor[HTML]{656565} 
\multicolumn{1}{|l|}{\cellcolor[HTML]{656565}{\color[HTML]{FFFFFF} }}                                & \cellcolor[HTML]{EFEFEF}Frequency: 4 GHz, ROB Size: 576, LSQ Size: 352, Issue Width: 8,   \\
\multicolumn{1}{|l|}{\multirow{-2}{*}{\cellcolor[HTML]{656565}{\color[HTML]{FFFFFF} \textbf{CPU}}}}  & Retire Width: 8, Branch Predictor: Perceptron                                             \\ \hline
\rowcolor[HTML]{9B9B9B} 
\multicolumn{1}{|l|}{\cellcolor[HTML]{9B9B9B}{\color[HTML]{FFFFFF} \textbf{L1D}}}                    & \cellcolor[HTML]{C0C0C0}Sets: 64, Ways: 12, MSHR Size: 16, Latency: 5 cycles              \\ \hline
\rowcolor[HTML]{656565} 
\multicolumn{1}{|l|}{\cellcolor[HTML]{656565}{\color[HTML]{FFFFFF} \textbf{L2C}}}                    & \cellcolor[HTML]{EFEFEF}Sets: 2048, Ways: 16, MSHR Size: 32, Latency: 10 cycles           \\ \hline
\rowcolor[HTML]{9B9B9B} 
\multicolumn{1}{|l|}{\cellcolor[HTML]{9B9B9B}{\color[HTML]{FFFFFF} \textbf{LLC}}}                    & \cellcolor[HTML]{C0C0C0}Sets: 4096/core, Ways: 12, MSHR Size: 64/core, Latency: 35 cycles \\ \hline
\rowcolor[HTML]{656565} 
\multicolumn{1}{|l|}{\cellcolor[HTML]{656565}{\color[HTML]{FFFFFF} }}                                & \cellcolor[HTML]{EFEFEF}MTPS: 4800 (LimitBW: 800), Channels: 1, Ranks: 1, Banks: 32       \\
\rowcolor[HTML]{656565} 
\multicolumn{1}{|l|}{\multirow{-2}{*}{\cellcolor[HTML]{656565}{\color[HTML]{FFFFFF} \textbf{DRAM}}}} & \cellcolor[HTML]{EFEFEF}tCAS: 15 ns, tRCD: 15 ns, tRP: 15 ns, tRAS: 32.5 ns               \\ \hline
\end{tabular}%
}
\caption{System Configuration}
\label{tab:system_config}
\end{table}
\section{State and Configuration}
SPPAM is designed for use alongside Berti in the L1D and Bingo in the LLC. Both designs follow what was originally proposed, but Bingo has been partitioned per-cpu and its overall state slightly reduced per-core to remain within state limits when 4 instances of Bingo are present. Berti utilizes an additional filtering table and provides stream-continuity information to SPPAM.
SPPAM's configuration is tabulated in Table~\ref{tab:sppam_config} and the state overheads of the prefetchers in each level of the cache are tabulated in Table~\ref{tab:sppam_state}.

\section{Evaluation Methodology}

We use the ChampSim-based \cite{Champsim} DPC4 simulator and provided traces for evaluating SPPAM. We evaluate against the baseline of DPC3 Berti in the L1D and Pythia in the L2C. We evaluate SPPAM's speedup in three different system configurations: single core, single core with limited bandwidth, and 4-core. The relevant details of each of these configurations are tabulated in Table~\ref{tab:system_config}.

For calculating the single-core speedups, the geometric mean of speedups across all traces is used. For multi-core, we calculate the harmonic mean of the speedup of the traces within each mix over the baseline prefetcher configuration.

\section{Experimental Results and Conclusion}
The speedups for each set of traces for both single core configurations are displayed in Figure~\ref{fig:singlecore_speedup}, with geomean speedups of 6.2\% and 5.9\% for standard and limited-bandwidth runs respectively. SPPAM offers substantial speedups for the sierra, bc, cc, sssp, gcc, mcf, and xalancbmk workloads over both the baseline and standard AMPM. Utilizing Bingo in the LLC offered slight performance improvements, with variations excluding Bingo seeing only a 5.5\% speedup and 4.9\% speedup for regular and bandwidth-limited runs.

Multi-core results are shown in Figure~\ref{fig:4core_speedup}, with each mix sorted from lowest to highest speedup and a geomean slowdown of 2.12\%. Bingo offers no performance advantages for SPPAM in multi-core, and could be excluded with minimal harm to the overall performance.
SPPAM fares well in single-core low-bandwidth simulations, suggesting that LLC and DRAM conflicts likely play a key role in the slowdown, rather than reduced resources in-and-of themselves. For these reasons, SPPAM's poor performance in multi-core can likely be resolved with proper tuning, and this will be pursued in future work.

\begin{acks}
The authors acknowledge the support from the Purdue Center for Secure Microelectronics Ecosystem [CSME\#210205].
\end{acks}

\bibliographystyle{ACM-Reference-Format}
\bibliography{references}

\appendix

\end{document}